# An Extended Hückel Theory based Atomistic Model for Graphene Nanoelectronics

HASSAN RAZA and EDWIN C. KAN

*School of Electrical and Computer Engineering, Cornell University, Ithaca, NY 14853 USA*

hr89@cornell.edu

**Abstract.** An atomistic model based on the spin-restricted extended Hückel theory (EHT) is presented for simulating electronic structure and I-V characteristics of graphene devices. The model is applied to zigzag and armchair graphene nano-ribbons (GNR) with and without hydrogen passivation, as well as for bilayer graphene. Further calculations are presented for electric fields in the nano-ribbon width direction and in the bilayer direction to show electronic structure modification. Finally, the EHT Hamiltonian and NEGF (Nonequilibrium Green's function) formalism are used for a paramagnetic zigzag GNR to show *$2e^2/h$* quantum conductance.

## 1. Introduction

Unconstrained graphene is a two dimensional (2D) semi-metallic material. The most important property that distinguishes it from other 2D materials/quasi-crystals is its linear instead of parabolic dispersion [1]. When graphene is patterned into a nanoscale ribbon by reducing one dimension, there are quantization effects dictated by *fixed* boundary conditions, which can result in a bandgap depending on the chirality and width. The most simplistic GNRs are zigzag GNR (zzGNR) and armchair GNR (acGNR) with zigzag and armchair edges, respectively. Conventionally, when a zzGNR is rolled into a carbon nanotube (CNT), it forms an armchair CNT and vice versa. Indeed in CNT, the quantization follows the *periodic* boundary conditions.Furthermore, by constraining one dimension, the dispersion of GNR is no longer linear and hence loses the inherent advantage of linear dispersion of graphene. Accurate device prediction demands an efficient model that can simulate realistic structures under different conditions of applied fields, boundary conditions in the form of attached molecules, multiple graphene layers, different substrates and hybrid structures with surrounding dielectric. A simple $p_z$ orbital tight binding (TB) model [1], although very efficient and useful for specific problems [2], cannot account for many necessary physical effects or the chemical nature of bonding. On the other hand, density-function-theory (DFT) based models are computationally prohibitive for systems having more than about 200 atoms.

Thus, a semi-empirical method such as the extended Hückel theory (EHT) seems to be a good tradeoff, since it is computationally inexpensive and captures most of the electronic and atomic-structure effects in the more rigorous methods. Using the EHT approach, systems with up to 1,000 atoms can be simulated easily [3]. For graphene, since EHT has been benchmarked with generalized gradient approximation (GGA) in DFT, it can provide accurate results that less sophisticated methods like local density approximation (LDA) may not capture resulting in underestimation of bandgap. Furthermore, we couple EHT with NEGF (Non-equilibrium Green's function formalism) for transport calculations.

## 2. Theoretical Model

We use spin-restricted EHT for the electronic structure calculations. The overlap matrix *S* is calculated from the well-defined non-orthogonal Slater

Type Orbital (STO) basis set of EHT, which is further used in calculating the Hamiltonian $H$. For C atom, we use the EHT parameters from [4] and are given in Table I. For electronic band-structure, the $H$ and $S$ matrices of the infinite GNR, graphene sheet and graphene bilayer is transformed to the reciprocal $k$-space as:

$$H(\vec{k}) = \sum_{n=1}^{N} H_{mn} e^{i\vec{k}\cdot(\vec{d_m}-\vec{d_n})} \qquad (1)$$

$$S(\vec{k}) = \sum_{n=1}^{N} S_{mn} e^{i\vec{k}\cdot(\vec{d_m}-\vec{d_n})} \qquad (2)$$

where $\vec{k}$ is the reciprocal lattice vector of the Brillouin zone and has 1D characteristics for GNRs and 2D for single and bilayer graphene sheets. The index $m$ represents the center unit cell and $n$ represents the neighboring unit cells, whereas $\vec{d_m}-\vec{d_n}$ is the relative displacement. The energy eigenvalue spectrum at a specific $k$ point is then computed. In order to calculate the band structure under the influence of an electric field $E$ in the width and bilayer directions, the Laplace potential $U_L(y)$ is included in the Hamiltonian:

$$U_L(y) = -eEy. \qquad (3)$$

Within the EHT scheme, $U_L$ is included as:

$$U_L(i,j) = \frac{1}{2}S(i,j)[U_L(y_i) + U_L(y_j)]. \qquad (4)$$

We also use an orthogonal TB scheme with hopping parameter $t$=3eV [1] for $p_z$ orbitals as comparison.

The NEGF formalism [5] is used for transport calculation. The Green's function with open electronic boundary conditions, in the form of the contact self-energies $\Sigma$'s, and electrostatic boundary conditions, in the form of $U_L$ defined by applied voltages, is given as:

$$G = [(E + i0^+)S - H - U_L - \Sigma_c]^{-1} \qquad (5)$$

The self-energy $\Sigma_c$ (where $c=1,2$) is defined as $[(E+i0^+)S_{dc}-H_{dc}]g_s[(E+i0^+)S_{cd}-H_{cd}]$, where the surface Green's function $g_s$ is calculated by a recursive technique. The contact broadening function, which has a physical meaning of carrier life time in the device region, is the non-Hermitian part of the self-energy, i.e. $\Gamma_c = i(\Sigma_c - \Sigma_c^+)$. Current is then given as:

$$I = \frac{2(\text{for spin})e}{h}\int_{-\infty}^{\infty} dE \cdot tr[\Gamma_1 G \Gamma_2 G^+](f_1 - f_2) \quad (7)$$

where $f_{1,2}$ are the Fermi functions of the two contacts.

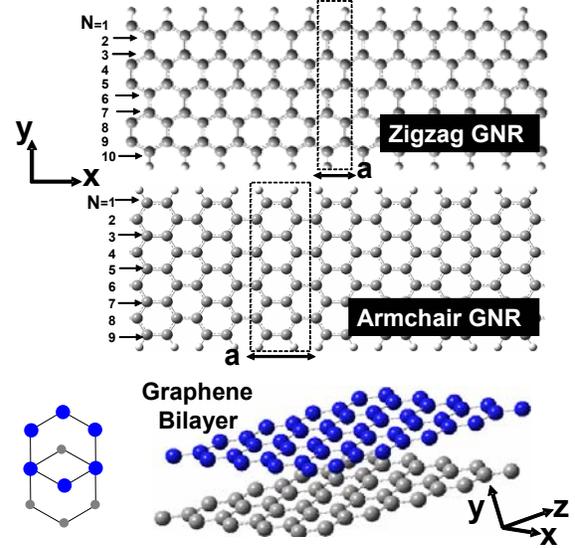

Figure 1. Visualization of zigzag graphene nanoribbon (zzGNR), armchair nanoribbon (acGNR) and bilayer graphene. The edges of zzGNR and acGNR are hydrogenated to eliminate surface states unless otherwise specified. A uniform electric field E can be applied in the width direction for GNRs and in the bilayer direction. The unit cells of GNRs are shown as dotted rectangles. Two benzene rings from top and bottom layer of the bilayer structure are shown to clarify that top layer is shifted in Bernal stacking. Visualizations are performed using GaussView [8].

In principle, the Hartree potential should also be solved and included in the Hamiltonian. However, for simplicity we do not consider it here and leave it for future work along with the spin-dependent EHT.

**3. Results and Discussions**

An $N$=10 zzGNR is shown in Fig. 1 and the band structures using EHT and TB schemes are shown in Fig. 2. Although there is much commonality between the two, the occupied bands deviate significantly and hence would lead to a different transmission response in this energy range. Figure 3 shows the band structures for $N$=9 acGNR, which again differ the most in the valence band and the deviation is more severe than in the case of zzGNR. Since the $sp^2$ bands are not present in TB calculations, the bandgap is over-estimated by 0.2eV. Further differences in dispersion appear at high energies. So, if the bandgap is corrected in TB models, they may be accurate for device modeling at operating voltages usually used. Moreover, TB has only the nearest neighbor coupling and hence any wave-function effects, in the form of second-nearest interaction that reflect the chemistry of the problem at hand, will be missing.

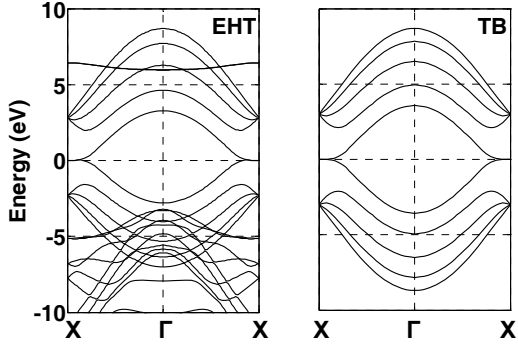

*Figure 2. Electronic structures of N=10 passivated zzGNR calculated using spin-restricted EHT and TB Hamiltonians. Around the Fermi energy, the dispersions look similar. However, in general, the E-k diagrams are quite different, in particular for the valence band states.*

Figure 4 shows the effect of an electric field in the width direction using the same acGNR. It is interesting to see that both EHT and TB give different dispersion as compared to Fig. 3. However, the bandgap remains the same. This has an interesting implication that the dispersions of a channel can be changed with a transverse field – an effect that may be utilized in future devices. Figure 5 shows the surface states due to dangling bonds for unpassivated zzGNR and acGNR [6] by EHT only, because the effect cannot be accounted for by TB. For both GNR, the dispersions in these bands are small as expected since these are localized on the unpassivated edges of GNRs. Furthermore, in acGNR, there is a finite density of surface states at the Fermi level, whereas for zzGNR, the surface states are below the Fermi level.

*Table 1*. EHT parameter for Carbon atom. $K_{EHT}$=2.80.

| Orbital | $E_{on-site}$ (eV) | $C_1$ | $C_2$ | $\zeta_1$ | $\zeta_2$ |
|---------|--------------------|-------|-------|-----------|-----------|
| 2s      | -20.315            | 0.740 |       | 2.037     |           |
| 2p      | -13.689            | 0.640 | 0.412 | 1.777     | 3.249     |

Figure 6 shows the electronic structure of a graphene bilayer and a graphene sheet. The bilayer has twice the bands with the degeneracy lifted due to finite coupling between the bands in the two graphene sheets separated by 3.35Å.

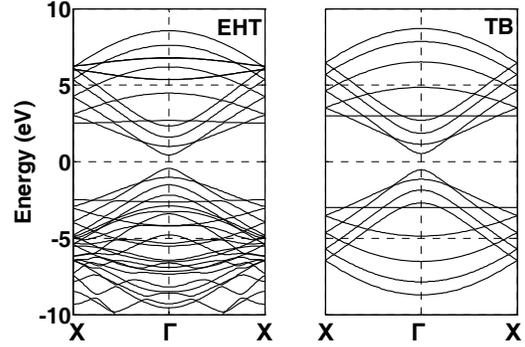

*Figure 3. Electronic structure of N=9 passivated acGNR by spin-restricted EHT and TB Hamiltonians. The differences are similar to those of zzGNR as in Fig. 2.*

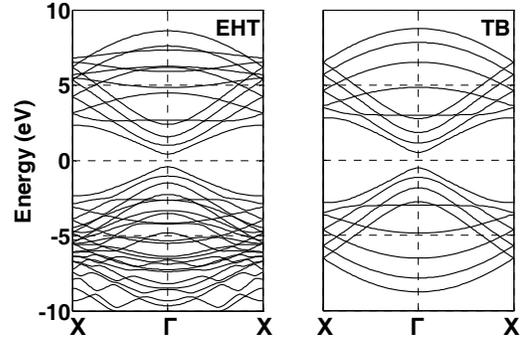

*Figure 4. Effect of electric field in the width direction on the electronic structure of N=9 passivated acGNR. By applying E=1V/nm in the width direction, the band gap roughly remains the same. However, the dispersions become different from Figs. 2 and 3.*

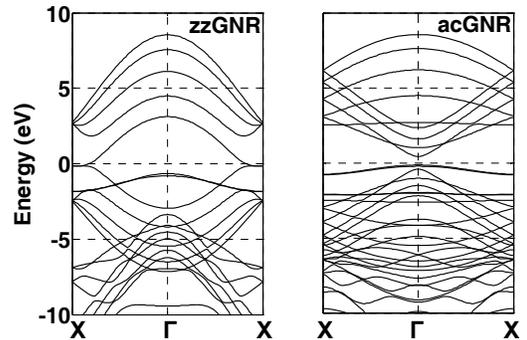

*Figure 5. Edge states for unpassivated zzGNR and acGNR, i.e. edges without H atoms. The dangling bonds introduce states below the Fermi energy for both configurations.*

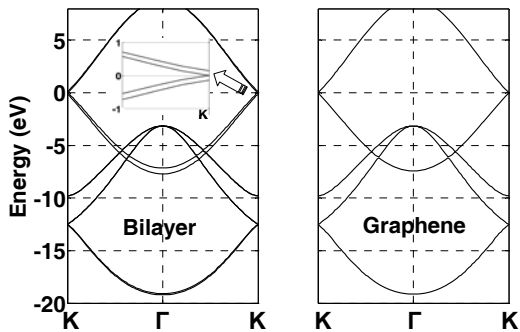

Figure 6. Comparison of electronic structure of a bilayer graphene and a graphene sheet. The degeneracy is lifted in the graphene bilayer due to overlapping $p_z$ orbitals of the two sheets. In the inset, a zoomed portion of E-k diagram is shown around K point to emphasize the lifting of degeneracy.

Figure 7 shows the effect of an electric field in the bilayer direction. A band gap is observed near the K edge of the Brillouin zone. However it is smaller than that in Ref. [7] probably due to the long-ranged basis set of EHT and/or not having Hartree potential in our calculations. The band edge shifts away from the K point with an increasing electric field. The resulting bandgap is also shown as a function of the electric field.

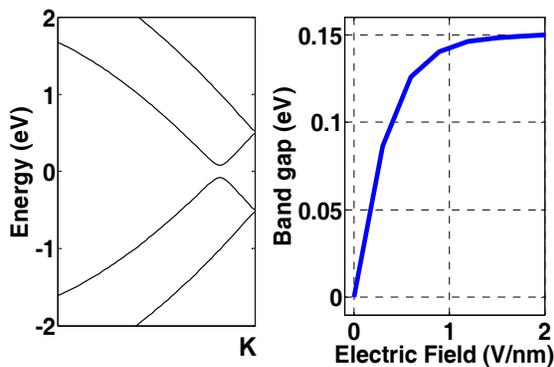

*Figure 7. Effect of the electric field in the bilayer direction, which results in a bandgap opening. The dispersion is shown around the K point for a 2V/nm electric field. The bandgap is proportional to the electric field until about 1V/nm and saturates afterwards.*

We finally present a transport calculation using the EHT and TB schemes in Fig. 8 for an N=10 intrinsic zzGNR with channel consisting of ten unit cells resulting in $L_{ch} \approx 2.5$nm. The source and drain contacts are assumed to be heavily doped GNRs and hence the drain voltage results in a linear Laplace potential drop inside the intrinsic channel. The source and drain equilibrium chemical potentials are placed at 10meV above intrinsic chemical potential. Due to near-zero bandgap and a finite voltage applied, band-to-band tunneling is expected to happen. For the paramagnetic phase considered here, a quantum conductance is observed. In one dimensional conductor, since the transport is independent of the dispersion of the channel, EHT and TB give the same result.

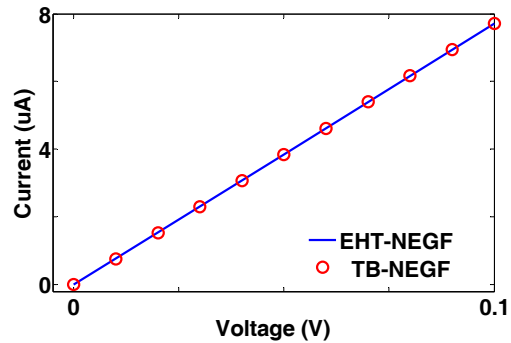

*Figure 8. Transport characteristics of N=10 paramagnetic zzGNR using NEGF. EHT and TB Hamiltonians give the same quantum of conductance, because in 1D transmission is independent of the band dispersion.*

## 4. Conclusions

We have put forward an EHT-based model that can be used for Graphene nanoelectronics overcoming many of the shortcomings of other electronic structure calculations and still providing an efficient platform for large heterogeneous devices with realistic structures. We apply our model to zzGNR, acGNR and graphene bilayer. Furthermore, we report the effect of the transverse electric field on the electronic structure of acGNR and graphene bilayer. A transport calculation of a paramagnetic N=10 zzGNR channel is shown to have $2e^2/h$ quantum conductance. The model can be extended to include spin-dependent and Hartree effects.

The work is supported by National Science Foundation (NSF) and by Nanoelectronics Research Institute (NRI) through Center for Nanoscale Systems (CNS) at Cornell University. We are grateful to Tehseen Raza for GaussView [8] visualizations.